\documentclass[journal,transmag]{IEEEtran}
\ifCLASSINFOpdf
\else
\fi
\usepackage{graphicx}
\usepackage{caption}
\usepackage{algorithm}
\usepackage{algorithmic}
\usepackage{amsmath}
\usepackage{balance}
\usepackage{amssymb}
\usepackage{amsmath}
\usepackage{colortbl}
\usepackage{booktabs}
\usepackage{multicol}
\usepackage{color}

\usepackage{multirow}

\begin{document}
%
\title{Duplicate matching and estimating features for detection of copy-move images forgery}


\author{\IEEEauthorblockN{Ghassem Alikhajeh \IEEEauthorrefmark{1},
Abdolreza Mirzaei\IEEEauthorrefmark{1},
Mehran Safayani\IEEEauthorrefmark{1}, and
Meysam Ghaffari \IEEEauthorrefmark{2} \\
ghasemalikhajeh@gmail.com, $\{$mirzaei,safayani$\}$@cc.iut.ac.ir , ghaffari@cs.fsu.edu
}
\IEEEauthorblockA{\IEEEauthorrefmark{1}School of Electrical and Computer Engineering,
Isfahan University of Technology, Isfahan , Iran}
\IEEEauthorblockA{\IEEEauthorrefmark{2}Department of Computer Science, Florida State University, Tallahassee, Florida ,32306 USA}
\thanks{Corresponding author: Meysam Ghaffari (email: ghaffari@cs.fsu.edu).}}

%



\IEEEtitleabstractindextext{%
\begin{abstract}
Copy-move forgery is the most popular and simplest image manipulation method. In this type of forgery, an area from the image copied, then after post processing such as rotation and scaling, placed on the destination. The goal of Copy-move forgery is to hide or duplicate one or more objects in the image. Key-point based Copy-move forgery detection methods have five main steps: preprocessing, feature extraction, matching, transform estimation and post processing that matching and transform estimation have important effect on the detection. More over the error could happens in some steps due to the noise. The existing methods process these steps separately and in case of having an error in a step, this error could be propagated to the following steps and affects the detection. To solve the above mentioned problem, in this paper the steps of the detection system interact with each other and if an error happens in a step, following steps are trying to detect and solve it. We formulate this interaction by defining and optimizing a cost function. This function includes matching and transform estimation steps. Then in an iterative procedure the steps are executed and in case of detecting error, the error will be corrected. The efficiency of the proposed method analyzed in diverse cases such as pixel image precision level on the simple forgery images, robustness to the rotation and scaling, detecting professional forgery images and the precision of the transformation matrix. The results indicate the better efficiency of the proposed method.
\end{abstract}

}

\maketitle

\IEEEdisplaynontitleabstractindextext

%
\IEEEpeerreviewmaketitle

\section{Introduction}
%
%
%
%
\IEEEPARstart{U}{sing} new image modification software, images could be easily modified. These images could be distributed easily over the internet\cite{birajdar2013digital}. So it is not easy to determine about the originality and accuracy of the images, especially if these images need to be used as an evidence in a court\cite{farid2009image}. To solve this problem, we need methods to detect the image modifications. These methods are known as image forgery detection methods\cite{christlein2012evaluation}.

Copy-move forgery is the simplest and most common digital image manipulation method. In this method, an area of the image will be copied and located in another area of the image\cite{amerini2011sift}. The goal of this method is to hide or duplicate one or few objects in the image\cite{nathalie2014survey}. Even amateurs could do this easily using image editing softwares such as Photoshop .Expert forger could modify the image such that the manipulation could not be detected by human eye, thus automatic methods are needed\cite{bayram2008survey}.

Detecting simple Copy-move forgery is easy, but in case of scaling or rotating the copied area, before locating in the image, detecting the forgery become hard\cite{thajeel2013state}. 

The proposed method is based on key points. In the previous methods, detection steps executed sequentially\cite{shivakumar2010detecting}, and in case of having an error in a step, this error propagated to the following steps which affects the detection. The goal of proposed method is to detect the errors and correct them by the interaction between the steps. The rest of this paper is as follows, an overview of the Copy-move forgery detection methods are described in Section 2. The proposed method is explained in section 3. The experimental results are in Section 4 and finally Section 5 is the conclusion.

\hfill 
 
\hfill

\section{Related Works}
Generally Copy-move forgery detection methods are categorized into two groups\cite{christlein2012evaluation}: block based methods and key point based methods. The block based methods have a blind search on the image, but the key point based methods detect the interesting points and key points in the image\cite{bayram2008survey}.

Forgery detection methods based on block have five major steps\cite{nathalie2014survey}: preprocessing, feature extraction, feature matching, pruning and post processing. These methods first apply needed preprocessing including converting the color image into grayscale on the image. Then divide the image into the overlapping square blocks. The appropriate features are extracted from each block. These features are extracted using diverse methods: 

(1) transform methods including DCT\cite{fridrich2003detection}\cite{huang2011improved}\cite{cao2012robust}, DWT\cite{zimba2011dwt}\cite{zhang2008new}\cite{bashar2010exploring}, Fourier-Mellin Transform (FMT)\cite{bayram2009efficient} and FT\cite{ketenci2014detection}.

 (2) Different color spaces: RGB\cite{luo2006robust}\cite{bravo2011exposing}, HSV\cite{liu2014hsv}, CMYK\cite{luo2006robust}\cite{bravo2011exposing} and gray scale luminance\cite{lynch2013efficient}.
 
 (3) texture such as local binary pattern\cite{alsawadi2013copy}.
 
 (4) Different moments such as blur moment invariants\cite{mahdian2007detection}, Zernike moment\cite{ryu2010detection}\cite{christlein2010rotation}. 
 
  (5) Dimensional reduction methods such as PCA\cite{popescu2004exposing}, KPCA\cite{bashar2010exploring} and SVD\cite{zhang2008new}\cite{kang2008identifying}.
  
  The extracted features from blocks need to be matched and similar blocks detected. In this step, blocks with similar block or blocks are detected. It is probable to have false matching, so in the next step false matches are pruned. Finally with methods such as morphology, small holes and the areas that increase the error are removed\cite{shivakumar2010detecting}.
  
  Key point based methods have five steps: preprocessing, feature extraction, key point matching, transform estimation and post processing. In these methods the key points are detected in the image and then features are extracted from these points. SURT\cite{mishra2013region} and SIFT\cite{amerini2011sift}\cite{huang2008detection}\cite{pan2010detecting}\cite{pan2010region} are well known methods in this category. Feature extraction based on MPEG-7 image signature tools has done in\cite{kakar2012exposing}.
  
  Zheng et al used Harris corner detector to detect key points and SURF descriptor to make feature vector for each key point\cite{zheng2014detection}. In \cite{zheng2012detection} key points are extracted based on local maximums or local minimums. They used angle and the distance ratio between each point and three nearest neighbors to extract features. Since rotation and scaling do not change the angle and distance ratio, these features are resistant to the rotation and scaling.
  
  In key point matching step, key point pairs are detected based on their feature vectors. In this step first, neighbor key points are detected based on feature vector and then, decide about matching between each key point with its neighbors. Sorting is the most used method for detecting the neighbors of each key point. Moreover, Best-Bin-First (BBF) method can detect nearest neighbors with low computations [29, 30].
  
  Analyzing the distance ratio between first nearest neighbor to the second nearest neighbor which called 2NN is a well-known matching method for the key points. The Euclidean [29, 31, 33] and inverse cosine angle of dot products [28] distance metrics are used. Another matching method is to compare the distance between each key point and its nearest neighbor with a threshold [30, 32, 34].
  
  2NN metric is not appropriate to detect forgery in images with multiple duplicated regions, because it just considers the two nearest neighbors for each key point. To solve this problem, Amerini proposed g2NN which has better results in detecting multiple duplicated regions in the image [4]. G2NN is a modification of the 2NN and considers the distance ratio of i th nearest neighbor to the i+1 th nearest neighbor. 
  
  In the estimation step, Affine transformation matrix is estimated and duplicated regions are detected. Affine transformation is estimated locally and generally. In the general mode, a transformation matrix is estimated for all key points such that this transform is compatible with the most of the points[30, 31]. In the local mode, multiple transformation matrix are estimated. In this mode the key points are clustered before the Affine transformation and then transformation matrix is calculated for each cluster[4, 28, 32].
  
  In [30] transform estimation is considered in three cases, copy-move, scale and rotation. In few methods, the cost function defined as transformation matrix error to estimate the Affine transform between matched pair points [4, 31]. Then the relationship of the transformation matrix estimation is calculated by optimizing the cost function. Furthermore, Random Sample Consensus (RANSAC)[4, 31] or Least Median of Squares (LMedS) [32] methods are used to improve the accuracy.
  
  To detect the duplicated area, transformation matrix is applied on the every pixel of the image and the correlation coefficient of each pixel and its transform is calculated[30-32]. Correlation coefficient is calculated by considering the luminance of each pixel and its neighbors with its transformed pixel. Then a threshold is used for this coefficient and the pixels with correlation coefficient higher than threshold are detected as duplicated. Finally morphology method is used to smooth the result.
  
  \section{Proposed Method}
  In key points based methods, transform estimation step(s) is the most important step(s). So each method with more precise estimated transformation matrix could has better results. In the all previous key points based methods, the method steps executed sequentially and independently. This means the following steps have no interaction with each other and just the output of each step is the input of the following step. Thus in case of having an error in one step, this error is propagated to the following steps and affects the final result. For example if noise or other reasons cause mis diagnosis in the matching step, this error affects the following step and detection system directly. Which means appropriate transformation matrix cannot be estimated from false key points. 
  
  Interaction between different system steps prevent these kind of errors. This interaction should be such that in case of happening error in a step, this error is detected, corrected and then process the following steps. Furthermore, the detection system must leave unchanged or strength the correct processes. For example in case of having error in matching step, the false matched pairs should be corrected and also correct matches must be strengthen.
  
  The error in the system steps can be detected by analyzing detection system strength. In the key points based methods having a more precise transformation matrix increase the detection precision, so the system strength can be calculated by the transformation matrix. 
  
  The most important steps in the detection system that need to be interacted are key points matching and transform estimation. In the proposed method, matching, clustering and transformation matrix estimation steps are interacted with each other to improve the system. This interaction is using a cost function. Table 1 defines the symbols that used in this paper.

  \begin {table}[H]
  \caption {Symbols} \label{tab:title} 
  \begin{center}
  \resizebox{\linewidth}{!}{
  \begin{tabular} {|p{1.2cm}|p{3cm}|p{3cm}|p{6cm}|}
  \hline 
  
   Variable &	Name	& Type	& Description  \\ 
  \hline  N &	Number of key points & 	Integer & 	Number of extracted points in the image \\ 
  \hline  $N_p$	& Number of matched points	& Integer	& Number of key points that paired in the matching step  \\ 
  \hline  C & 	Number of Clusters	& Integer	& Number of clusters in the clustering algorithm   \\ 
  \hline  dim & 	Dimensions of each point &	Integer	& Number of dimensions of each point. Here dim=3 and each point is represented by (x,y,1) vector  \\ 
  \hline  K	& Number of nearest neighbors	& Integer	& For each point such as $X_k$, maximum number of nearest neighbors (in the descriptor space) that can match with $X_k$  \\ 
  \hline  P	&  Matching coefficient	& Real	& This coefficient determines the effect of points error in the key points matching matrix. greater matching coefficient indicates that greater errors have more matching and vise versa. \\ 
  \hline  U	& Segmentation matrix	& Matrix with size $(C*N_p)$	& This matrix shows the dependability of each point to each cluster C. its value is in the range of [0 1] and known as membership value. In this matrix, Uik shows the membership value of key point k in the cluster i \\ 
  \hline  m	& Fuzzification coefficient	& Real	& This coefficient determines the shape of membership function in clustering (m>1). Values close to 1 make the clustering algorithm crisper. Increasing m make the algorithm (and membership function) more fuzzy \cite{ghaffari2016ambiguity}\cite{ghadiri2016bigfcm}. \\ 
  \hline X	& Key points matrix	& A matrix with N*dim dimension	& This matrix includes the spatial coordination of the key points that detected in the feature extraction step (each row is a key point). \\ 
  \hline  V	& Cluster centers matrix	& A matrix with the size of C*dim	& This matrix contains the extracted centers from the clustering algorithm \\ 
  \hline  Des	& Descriptors matrix	& A matrix with the size: N*128 	& This matrix contains the key points descriptor that extracted from SIFT algorithm  \\ 
  \hline  $\alpha$	& Key points matching matrix	& A matrix with the size:
  $N_p*N$	& This matrix contains the value of matching between each key point with its neighbors
   \\ 
  \hline  $H^i$	& i th cluster transformation matrix	& A matrix with the size:
  3*3& 	$H^i$ $(1\leq i \leq C)$ shows the transformation matrix of the i th cluster
   \\ 
  \hline 
  \end{tabular} 
  }
 
  \end{center}
  \end {table}

  \subsection{cost function}
  The goal of this function is clustering and matching the data such that the error of the transformation matrix and key points clustering is minimized. A simple cost function could be defined as:
  
  \begin{multline}
  Q=\Sigma_{i=1}^C \Sigma_{k=1}^{N_p}U_{ik}^m \{\parallel x_k - v_i \parallel ^2 +  \\ \Sigma_{x_{k'}\in K \space Nearest \space Neighbor \space of \space x_k \alpha_{kk'}^p  \parallel Des_{x_k}^2 - Des_{x_{k'}} ^2 } \parallel \parallel x_{k'} - H^ix_k \parallel^2  \} \\
  \\
  \Sigma_{i=1}^C U_{ik}=1, k=1,2,\dots, N_p \\
  \\
  \Sigma_{x_{k'}\in K_Nearest Neighbor of x_k} \alpha_{kk'} = 1 , k=1,2,\dots, N_p
  \end{multline}
  
 
 Where Des, K, m, U, V, X, $H^i$, $\alpha$, N, $N_p$, C, P are introduced in table 1 and $\parallel X_k’-H^iX_k\parallel$ is the transformation matrix error between $X_k$ and $X_k’$ key points. $\parallel Des_{x_k}-Des_{x_k'}\parallel$ is the distance of the descriptor between $X_k$ and $X_k’$ key points. Furthermore, the fitness function has two constraints. These constraints will be added to the cost function in the following sections using the Lagrange coefficient. The symbol $\parallel.\parallel$ is the Euclidean norm. For example the Euclidean norm of vectors $X_k$  and $V_i$ is defined as:
 
 \begin{equation}
 \parallel x_k- v_i \parallel^2=\Sigma_{j=1}^{dim}\{   x_{kj}-v_{ij} \} ^2 
 \end{equation}
 
 First term in the cost function (equation 1) tries to minimize the spatial distance between key points and the centers and the second term minimize the key points transform error with their \textit{K} nearest neighbors error by considering the matching value and their descriptors. This means the goal of first term is optimum clustering and the goal of second term is to estimate the optimum transformation matrix. In the equation 1, for each key point \textit{k}, its membership value in each cluster, its distance from the cluster center and transformation matrix error are affect the cost function and its sum for all of the key points is defined as the cost function.
 The U matrix in equation (1) shows the membership values of each key point in different clusters. The descriptor matrix (Des) includes the key points that $Des_{x_k}$  shows the descriptor of $x_k$ key point. The goal of using key points descriptor distance in the cost function is to prioritize the nearest neighbors for each key point which means nearest neighbors has more effect than far neighbors.
 In equation (1), each key point affects the error calculation by considering the matching value and the closeness value (in the feature space). To prevent slow down and increase the complexity of the algorithm, instead of using all nearest neighbors just \textit{K} nearest neighbors has been used.

Key points matching matrix (α) shows the matching values of each key point with it’s K nearest neighbors. This matrix is initialized in the key points matching step. The initialization is such that for each key point, the first nearest neighbor has the value of 1 and K-1 nearest neighbors has the value of 0. Then in the algorithm process for each key point in the Α matrix and the values of K nearest neighbors are updated and get the value in range of [0 1]. The bigger value means the stronger matching.

Using the $\alpha$ matrix make the algorithm robust against false matching. These false matching could be due to the noise. So the $\alpha$ matrix make the algorithm more flexible and in case of false matching, it can be detected and corrected. By analyzing the matching error, false matches can be detected.  
 
For the correction using matching with one of the K nearest neighbors instead of all nearest neighbors is possible. In this case the neighbors of the falsely matched key points are used to correct the matching. So if matching of a point like X with its nearest neighbor is false, the error is increased and the matching value in the Α matrix of the algorithm must be decreased. Moreover, second to K th nearest neighbors could be better matches for the X point. In this case the value of best match must be increased in the Α matrix. This means that the $\alpha$ matrix lets us to choose the best matched point from the k nearest neighbors. In this matrix most of the matches may remain unchanged and the algorithm matches them correctly in the beginning.

The transformation matrix between key points in the i th cluster is showed by $H^i$. The main objective of the proposed method is to achieve the transformation matrixes with minimum error because these transforms are used to detect and determine the location of the copied areas. Each affine matrix $H^i$ is defined as:

\[
H^i=
  \begin{bmatrix}
    h^i_{11} &  h^i_{12} &  h^i_{13} \\
   h^i_{21} &  h^i_{22} &  h^i_{23} \\
   h^i_{31} &  h^i_{32} &  h^i_{33}
  \end{bmatrix} 
  =
  \begin{bmatrix}
      h^i_{11} &  h^i_{12} &  h^i_{13} \\
     h^i_{21} &  h^i_{22} &  h^i_{23} \\
     0 &  0 &  1
    \end{bmatrix}
    =
  \begin{bmatrix}
        A &  t \\
       0^T &  1       
      \end{bmatrix}  
\]

Where $h_{pq}^i$ is the p th row and q th column of transformation matrix in i th cluster and A matrix is the representation matrix of rotation and scaling. t shows the move vector between key points [32]. So the Affine transformation matrix is defined from rotation and scaling with matrix A and moving with vector t.

The most important part of equation 1 is the transformation matrix error $\parallel x_{k'}-H^i x_k\parallel $ which defined using geometric error. In the next subsection, this error will be defined and the calculation method of V, U, Α, $H^i$ matrixes using this error will be explained.

\subsubsection{cost function based on the geometric error}
The geometric error of the transformation matrix is defined as[35]:

\begin{equation}
\parallel x_{k'} -H^i x_k \parallel = \Sigma_{m=1}^{dim} (x_{k' m}-\Sigma_{n=1}^{dim}h_{mn}^i x_{kn})^2 
\end{equation}

Thus the final cost function based on geometric error is by using equation 2 and 3 in equation 1 and defined as:

\begin{multline}
Q=\Sigma_{i=1}^C \Sigma_{k=1}^{N_p}u_{ik}^m \{ \Sigma_{j=1}^{dim}(x_{kj}-v_{ij})^2 +  \\
\Sigma_{x_{k'} \in K-Nearest Neighbor of x_k} \alpha_{kk'}^P \parallel Des_{x_k}-Des_{x_{k'}}  \parallel ^2 \\  \Sigma_{m=1}^{dim}(x_{k' m}-\Sigma_{n=1}^{dim}h_{mn}^i x_{kn})^2 \}
\end{multline}

Which can be simplified as:

\begin{multline}
Q= \\ \Sigma_{i=1}^C \Sigma_{k=1}^{N_p}u_{ik}^m \{ LD_{ki} +\Sigma_{x_k' \in KNN  of x_k} \alpha_{kk'} ^P DD_{kk'}\parallel x_{k'}-H_{x_k}^i \parallel \} \\
DD_{kk'} = \parallel Des_{x_k}-Des_{x_k'}\parallel^2 \\
LD_{ki} = \Sigma_{j=1}^{dim}(x_{kj}-v_{ij})^2 
\end{multline}

That $LD_{ik}$ is the spatial distance of the key point k with the center i. $DD_{kk’}$ is the descriptor distance of xk,xk’ point. In this equation X and Des matrixes are constant and $\alpha$,U,V, $H_i$ must be calculated. To optimize the cost function of the equation 5 and calculating U,V,$H_i$, $\alpha$ gradient is used and the following equations are calculated:
\begin{multline*}
\frac{\sigma Q}{\sigma u_{pq}}=0,     p=1,2,…,C    ,    q=1,2,…,N_p \\
\frac{\sigma Q}{\sigma v_{pq}}=0,     p=1,2,…,C    ,    q=1,2,…,dim \\
\frac{\sigma Q}{\sigma \alpha_{pq}}=0,     p=1,2,…,N_p     ,    q=1,2,…,N \\
\frac{\sigma Q}{\sigma h_{pq}^i}=0,     p=1,2    ,    q=1,2,3
\end{multline*}

That p and q are the desired row and column. First the derivation of the Q based on $V_{pq}$ is calculated and set it equal to zero. $V_{pq}$ is the q th dimension of the p th cluster center.

\begin{equation}
\frac{\sigma Q}{\sigma v_{pq}} = \Sigma_{k=1}^{N_p}u_{pk}^m(x_{kq}-v_{pq})=0
\end{equation}

Finally the cluster centers are calculated by:

$
v_{pq}=\frac{\Sigma_{k=1}^{N_p}u_{pk}^m x_{kq}} {\Sigma_{k=1}^{N_p}u_{pk}^m}
$

Derivation of the cost function based on the Affine transformation matrix elements is calculated as equation 7:

\begin{multline}
\frac{\sigma Q}{\sigma h_{pq}^i} = \Sigma_{k=1}^{N_p}u_{ik}^m \{ \Sigma_{k'\in KNN} \alpha_{kk'}^P DD_{kk'} \\ (-2x_{kq} (x_{k' p}-\Sigma_{n=1}^{dim}h_{pn}^i x_{kn})) \} \\ =-2\Sigma_{k=1}^{N_p}u_{ik}^m \Sigma_{k'\in KNN}\alpha_{kk'}^P DD_{kk'} x_{kq} x_{k'p} \\ +2\Sigma_{k=1}^{N_p}u_{ik}^m \Sigma_{k'\in KNN} \alpha_{kk'}^P DD_{kk'} x_{kq} \\ \Sigma_{n=1}^{dim}h_{pn}^i x_{kn}=0
\end{multline}

The equation 7 is calculated for the all elements of the $H^i$ and for each element there is an equation with three unknowns. For each row, three elements of it are exist in the equation. So two linear equations with three obviouses and three unknowns are extracted. By solving these equations, the values of the Affine transformation matrix are extracted.

The constraints of the matrix U must be added to the cost function before calculating it. In the U matrix, sum of the membership values of each point to the all clusters must equal to one, which means:

\begin{equation}
\Sigma_{i=1}^C u_{ik}=1,       k=1,2,…,N_p
\end{equation}
To apply this constraint to the cost function, Lagrange multipliers are used. So the cost function with constraint for each key point is as follows:
\begin{multline*}
Q_2=\Sigma_{i=1}^Cu_{ik}^m \{LD_{ki}+\Sigma_{x_{k'} \in KNN  of x_k} \\ \alpha_{kk'}^P DD_{kk'} \parallel x_{k'}-H^i x_k \parallel \} +\lambda(\Sigma_{i=1}^C u_{ik} -1) \\
  k=1,2,…,N_p
\end{multline*}

That $\lambda$ is the Lagrange multiplier. So the following derivations must be equal to zero:
\begin{multline*}
\frac{\sigma Q_2}{\sigma u_{pq}}=0,      p=1,2,\dots,C    $\space$ , $\space \space \space$    q=1,2,\dots,N_p \\
\frac{\sigma Q_2}{\sigma \lambda}=0
\end{multline*}

The derivation of the cost function $Q_2$ based on the $\lambda$ multiplier shows the constraint and the derivation based on the p and q elements of the U matrix is as:
\begin{multline}
\frac{\sigma Q_2}{\sigma u_{pq}}=mu_{pq}^{m-1} ( LD_{qp} +\Sigma_{x_{k'} \in KNN  of x_q} \alpha_{qk' }^P DD_{qk'} \\  \parallel x_{k'}-H^p x_q \parallel )+\lambda  \\
So:\\
u_{pq}=\frac{-\lambda}{m}^{\frac{1}{m-1}} \\ ( \frac{1}{LD_{qp} +\Sigma_{x_{k'} \in KNN  of x_q} \alpha_{qk'}^P DD_{qk'} \parallel x_{k'}-H^p x_q \parallel)^{\frac{1}{m-1}}}
\end{multline}

By applying normalization $\Sigma_{j=1}^C u_{jk}=1$ in the equation 9, we have:

\begin{multline}
(\frac{-\lambda}{m})^{\frac{1}{m-1}} \\ (\Sigma_{j=1}^C \frac{1}{(LD_{qj} +\Sigma_{x_{k'}\in KNN  of x_q} \alpha_{qk'}^P DD_{qk'} \parallel x_{k'}-H^j x_q \parallel )^\frac{1}{m-1} }) \\ =1 \\
Thus: \\
(\frac{-\lambda}{m})^{\frac{1}{m-1}} = \\ \frac{1}{(\Sigma_{j=1}^C \frac{1}{(LD_{qj} +\Sigma_{x_{k'}\in KNN  of x_q} \alpha_{qk'}^P DD_{qk'} \parallel x_{k'}-H^j x_q \parallel )^\frac{1}{m-1} })} 
\end{multline}
And finally by replacing the equation 10 in the equation 9, the final equation for the clustering matrix of the cluster p and key point q is as equation 11.
\begin{multline}
U_{pq} = \\ \frac{1}{ \Sigma_{j=1}^C ( \frac{(LD_{qp} +\Sigma_{x_{k'}\in KNN  of x_q} \alpha_{qk'}^P DD_{qk'} \parallel x_{k'}-H^p x_q \parallel )}{(LD_{qj} +\Sigma_{x_{k'}\in KNN  of x_q} \alpha_{qk'}^P DD_{qk'} \parallel x_{k'}-H^j x_q \parallel ) }) ^\frac{1}{m-1} } 
\end{multline}

Based on the equation 11, if a point such as p is close to a specific center such as q and its transform error with transformation matrix $H^p$ is low, the value of $U_{pq}$ is large and vice versa. Which means the membership value of the key points in the clusters has inverse ratio with the distance from the center and the transformation matrix error.

In the $\alpha$ matrix, each point has constraint. In this matrix the sum of the matching value of each point with its K nearest neighbors is 1:

\begin{equation}
\Sigma_{k'\in KNN of x_k} \alpha_{kk'} =1,       k=1,2,\dots,N_p
\end{equation}

This constraint is added to the cost function using the Lagrange multipliers same as clustering matrix. The cost function with constraint for each key point is as follows:

\begin{multline*}
Q_3=\Sigma_{i=1}^C u_{ik}^m \{ LD_{ki} +\Sigma_{x_{k'} \in KNN  of x_k} \alpha_{kk'}^P DD_{kk'} \\  \parallel x_{k'}-H^i x_k \parallel \} + \lambda(\Sigma_{x_{k'} \in KNN  of x_k} \alpha_{kk'} -1 ) \\
k=1,2,\dots,N_p
\end{multline*}
For calculating the Α matrix, the following derivations must be calculated:
\begin{multline*}
\frac{\sigma Q_3}{\sigma \alpha_{pq}}=0,      p=1,2,\dots,N_p     ,     q=1,2,\dots,N \\
\frac{\sigma Q_3}{\sigma \lambda}=0
\end{multline*}

The second derivation shows the constraint. The derivation based on the Α matrix elements routine is like the clustering matrix elements derivation.

\begin{multline}
\frac{\sigma Q_3}{\Sigma \alpha_{pq}}=\Sigma_{i=1}^C u_{ip}^m  ( P\alpha_{pq}^{P-1} DD_{pq} \parallel x_q-H^i x_p \parallel)+\lambda \\
So: \\
\alpha_{pq}=(\frac{-\lambda}{P})^{(\frac{1}{P-1})}  (\frac{1}{(\Sigma_{i=1}^C u_{ip}^m DD_{pq} \parallel x_q - H^i x_p \parallel )^{(\frac{1}{P-1})}} )
\end{multline}
By applying the normalization $\Sigma_{(x_{k'}\in KNN  of x_k} \alpha_{kk'} =1 $ in the equation 13, the following equation is derived:

\begin{multline}
(\frac{\lambda}{P})^{(\frac{1}{P-1})} \\  ( \Sigma_{x_{k'} \in KNN  of x_p} \frac{1}{(\Sigma_{i=1}^C u_{ip}^m DD_{pk'} \parallel x_{k'} -H^i x_p \parallel )^{\frac{1}{P-1}}} )=1 \\
Thus: \\
(\frac{-\lambda}{P})^{\frac{1}{P-1}}= \\ \frac{1}{(\Sigma_{x_{k'} \in KNN  of x_p} \frac{1}{(\Sigma_{i=1}^C U_{ip}^m DD_{pk'}\parallel x_{k'}-H^i x_p \parallel )^{\frac{1}{P-1}} } )}
\end{multline}

Finally the key points matching matrix update equation is as equation 15:

\begin{equation}
\alpha_{pq}= \frac{1}{(\Sigma_{x_{k'} \in KNN of x_p} (\frac{\Sigma_{i=1}^C U_{ip}^m DD_{pq} \parallel x_q - H^ix_p \parallel}{\Sigma_{i=1}^C U_{ip}^m DD_{pk'} \parallel x_{k'} -H^i x_p \parallel }) ^{\frac{1}{P-1}}   )}
\end{equation}

Figure 1 shows the effect of using matching and clustering matrixes. In this figure, the area A is copied twice and the copied areas are showed with B and C symbols. In this figure, the key points of the area A are clustered into two clusters that shown by blue and red. In the area A, a key point is showed by a1 that its membership value to the red cluster is high, but its matching is consistent with the key points in the blue cluster. In this case if one of its second to k th nearest neighbors of the key point a1 are in the area C, then in the matching matrix, the matching value of the key point a1 with its nearest neighbor is decreased and the matching value with the nearest neighbor in the area C is increased (the reason is that the error of the key point a1 with first nearest neighbor is high and the error with the nearest neighbor in the area C is low (equation 15)). In case of none of the second to K th neighbors are not exist in the area C, then the membership value of the key point a1 in the red cluster is decreased and its membership value in the blue cluster is increased. The reason is that the distance of the key point a1 with the centers of the blue and red cluster are close but the transform error in the blue cluster is high and in the red cluster is low (equation 11).

\begin{figure}
\centering
\includegraphics[width=0.7\linewidth]{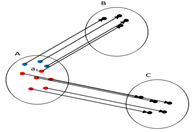}
\caption{The effect of the clustering and matches matrixes}
\label{fig:1}
\end{figure}

\subsection{steps of the proposed method}
The proposed method has six steps: preprocessing, feature extraction, basic key points matching, transformation matrix estimation, area detection and post processing.

\subsubsection{preprocessing}  In this step, the RGB Images are converted to the bit maps.

\subsubsection{feature extraction}
In this step, first the key points are detected and then for each points, the descriptors are made based on the SIFT algorithm[36].

\subsubsection{basic key points matching}
in this step, the matched pairs of points are extracted from the detected key points. Matching is based on the g2NN metric so the algorithm can detect areas that copied multiple times. The g2NN metric is defined with the ratio $d_i/d_(i+1)$ where $d_i$ is the Euclidean distance with i th nearest neighbor $1\leq i\leq N$. if this ratio is lower than a defined threshold, two key points are matched.

Choosing optimal threshold is important. Small threshold cause few matched points. In this case some of the really paired points could not be detected. By having a large threshold, so many matched points are detected and some of them may detected falsely. In the previous methods the 0.5 is considered as the threshold which cause not to detect some of real matching and moreover, matches are not detected in small areas.

In the proposed method, to preserve the true matches, the threshold is equal to 0.7. Using this value, the algorithm finds more matches and most of the true matches are detected. In this case, it is probable to have some false detections, which will be eliminated in the following steps. Thus the matches are distributed along the copied area and matched key points are detected for the small areas. The Des matrix is calculated in this step and Α matrix is initialized.

\subsubsection{transformation matrix estimation}
The Affine transformation matrixes is estimated for each cluster in this step. At the beginning the clustering matrix must be initialized. For this work, two methods are available: random initialization and initialization using Fuzzy c-means. The random initialization method is like blind search and decrease the speed and efficiency of the algorithm. Thus Fuzzy c-means method is used in the proposed method for the initialization.

After the initializing U matrix, the centers of the samples are calculated using equation 6. Then Affine transformation matrix is estimated for every cluster using the equation 7. Then Α matrix must be updated to decrease the effect of the false matches in error calculation (equation 15). Finally U matrix is updated using transformation matrix error and the points distance from the centers (equation 11). Then the error is calculated for each key point and points that have greater error than the threshold are removed. The threshold starts with a large value and decrease in iterations as showed in equation 16:

\begin{equation}
T=T_{max} - \frac{1}{1+e^{\frac{-1*(\frac{iter}{iter_{max}}-\theta)}{\tau}}}
\end{equation}

Where $T_{max}$ is the maximum threshold, $T_{min}$ is the minimum threshod, iter is the current iteration, $iter_{max}$ is the maximum iteration number, $\tau$ and $\theta$ are controlling decrease value. Lower $\tau$ and $\theta$ values cause great decrease in the prime iterations and smooth decrease in the last iterations. In the experiments, $\theta$=0.001 and $\tau$ =0.12 is used to have small threshold changes in the last iterations. Furthermore, $T_{max}$=2000 and $T_{min}$=0.1 is used. Figure 2 shows the threshold changes.
\begin{figure}
\centering
\includegraphics[width=0.7\linewidth]{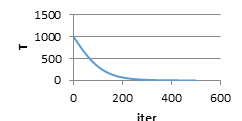}
\caption{Threshold changes in iterations}
\label{fig:2}
\end{figure}

V, $H^i$, Α and U matrixes are updated iterMax times. In the experiments iterMax=500 is used. To improve the transformation matrix, in another routine, the transformation matrix is estimated (like the above routine). The $\alpha$ matrix is considered constant at the beginning and each key point is matched with one of its K nearest neighbors that have the greatest value in the $\alpha$ matrix. Then the above iterative routine is executed and the transformation matrix is estimated again.

\subsubsection{detecting duplicated regions}
After executing the previous step, the Hi transformation matrixes are estimated for all clusters which used to detect duplicated regions. For each pixel in the image, the transformation matrix is applied on the pixel. Then a 7*7 window around the pixel is considered which showed by A. For that pixel, the transformation matrix of the nearest cluster to the pixel (by calculating the distance between pixel and the centers of the clusters) is selected and the value of rotation, scaling and moving is extracted. Rotation and scaling is applied on the area A (the window around the pixel) and the new area is called Ars. Then an area with the size of Ars is selected around the transformed pixel which called B. Figure 3 shows the routine of selecting these areas for a sample pixel. Finally the correlation coefficient between Ars and B areas is calculated. This coefficient is defined as equation 17:

\begin{equation}
Corr(x)=\frac{\Sigma_{a \in A_{rs}}\Sigma_{b \in B} (I(a)-\mu_{A_{rs}})*(I(b)-\mu_B)}{\sqrt{(\Sigma_{a \in A_{rs}}(I(a)-\mu_{A_{rs}})^2)*(\Sigma_{b \in B }(I(b)-\mu_B)^2)}}
\end{equation}

Where I(x) is the luminance of the pixel x in the image, $\mu_{A_{rs}}$  is the average luminance of the pixels in the area $A_{rs}$, $\mu_B$ is the average luminance of the pixels in the area B. this coefficient is calculated for each pixel in the image and the result is a matrix with the equal size to the original image. The value of this coefficient is in range of [-1 1]. After calculating Corr matrix for every pixels of the image, this matrix must converted to the binary. This work is done by using a threshold. This threshold determines the output value and its large value decrease the error and if its value is low, the error will increase. In the experiments 0.6 is used for this threshold.

\begin{figure}
\centering
\includegraphics[width=0.7\linewidth]{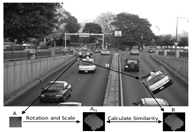}
\caption{Applying transform matrix on a sample pixel}
\label{fig:3}
\end{figure}

\subsubsection{Post Processing}
in this step, the areas which their size is below $0.1\%$ of the image are removed. And Morphology is used to soften the areas. Algorithm 1 shows the Pseudo code of the proposed method.

     \begin{algorithm}
      \caption{The pseudo code of the proposed method}
    \begin{algorithmic} 
     \label{alg:Pseudomodel}
   \STATE ICMFD (input image, C, P, K, m, $T_{max}, T_{min}$, $iter_{max}$)
   \STATE 1)	Preprocessing: convert the image to gray scale.  
    \STATE 	2)	Feature extraction using SIFT: \\
    	a.	Scan the image for keypoints \\
    	b.	Calculate a feature vector $\vec{f}_i$ for every keypoint
    \STATE 3)	Matching Feature: \\
    a.	Find matched keypoints with g2NN  \\
    b.	Initialize $\alpha$. $\alpha_{ij}=1$, if keypoint i matched with j and $\alpha_{ij}=0$ otherwise
    \STATE 4)	Initialize U using fuzzy c-means clustering
    \STATE 5)	Estimate transformation  \\
    a.	iter = 1  \\
    b.	Calculate V using 3-6 \\
    c.	Calculate $H_i (1\leq i\leq C)$ using solving two systems of equations 3-7 \\
    d.	Calculate $\alpha$ using 3-15  \\
    e.	Calculate U using 3-11 \\
    f.	Calculate Threshold T using 3-16 \\
    g.	Calculate transformation error for each keypoint and remove keypoint with error $>$ T \\
    h.	If $(iter < iter_{max})$ then iter = iter + 1 and goto step b, else goto step 7
    
    \STATE 6)	Fix $\alpha$
    \STATE 7)	Estimate transformation  \\
    a.	iter=1   \\
    b.	Calculate V (equ 3-6), $H_i (1\leq i\leq C)$ (equ 3-7) and U (equ 3-15) \\
    c.	If $(iter < iter_{max})$ then iter = iter + 1 and goto step b, else goto step 9 
    
    \STATE 8)	Regions detection  \\
    a.	For each pixel in image apply H of nearest cluster on them and calculate correlation coefficient (equ 3-17) for that pixel. Correlation coefficients are saved in Corr that have the same size with input image.\\
    b.	Binerization Corr with the threshold 0.6.
    
    \STATE 9)	Post processing: fill regions less than 0.1 of size total image and use morphological operations to smooth regions. 
    
    \end{algorithmic}
    \end{algorithm}    
    
\section{Experimental Results}
The efficiency of the proposed method is analyzed in this section. To compare the detection precision, two datasets from [4] are used. These datasets include original and manipulated images and images in the second dataset have higher resolution. The manipulated images made by various attack scenarios including different combination of scaling and rotation. In the second dataset the various attack scenarios has been covered but in the first dataset few attacks has been applied. So we generate manipulated images in the first dataset with other attack scenarios which shown in table 2. This table includes 15 different attacks and each attack, the value of rotation and scaling in x,y scale is shown with $\theta$, $s_x$,$s_y$. this table covers different attack scenarios including symmetric scaling, asymmetric scaling, scaling up and down and different combinations of rotation and scaling.

The first dataset includes 110 original images and 150 manipulated images. The 150 manipulated images are generated by applying 15 attack scenarios on 10 images (table 2). Each manipulated image is generated by choosing a square or rectangular region and applying the intended attack scenario. So the first dataset includes 260 images and the second dataset includes 2000 images which has 1300 original images and 700 manipulated images.

\begin {table}[H]
\caption {Different attack scenarios} \label{tab:title} 
\begin{center}
\begin{tabular} {|c|c|c|c|} \hline
  \centering{\textbf{Attack}} & \textbf{$S_x$} & \textbf{$S_y$} & \textbf{$\theta$} \\ \rowcolor[gray]{.9}
       \hline
      A1 & 	1	& 1	& 0      \\ 
       \hline
     A2	 & 1	& 1	& 10 \\  \rowcolor[gray]{.9}
       \hline
      A3 & 	1 & 	1	& 20 \\
       \hline
      A4	& 1	& 1	& 30 \\ \rowcolor[gray]{.9}
        \hline
       A5 & 	1 & 	1 & 	40  \\
          \hline
        A6	& 1	 & 1	& 50 \\ 
        \rowcolor[gray]{.9}
                  \hline    
      A7	& 1.2	& 1.2	& 0 \\ 
                         \hline    
      A8 & 	1.3 & 	1.3	& 0 \\ 
      \rowcolor[gray]{.9}
                         \hline    
      A9	& 0.8	& 0.8 & 	0 \\ 
                        \hline    
      A10 & 	0.75	& 0.85 & 	0 \\
      \rowcolor[gray]{.9} 
                        \hline    
     A11	& 0.85	& 0.75	& 0 \\ 
                       \hline    
    A12	& 1.2	& 1.2	& 30 \\ 
    \rowcolor[gray]{.9}
                       \hline    
    A13	& 0.8	& 0.8	& 30 \\ 
                       \hline   
     A14	& 0.75	& 0.85	& 35 \\ 
     \rowcolor[gray]{.9}
                       \hline    
    A15	& 1.4	& 1.2	& 35 \\ 
                       \hline           
\end {tabular}
\end{center}
\end {table}

To analyze the efficiency of the algorithm and comparing it with other methods, the false positive and true negative metrics are used which defined as:

\begin{multline*}
TPR=\frac{TP}{TP+FN}\\
FPR=\frac{FP}{FP+TN}\\
\end{multline*}
 
 Where TP is the true positive, FP is the false positive, TN is true negative and FN is false negative. In image detection mode, TP metric shows the number of the manipulated images that detected truly, FP is the number of original images that falsely detected as manipulated, TN is the number of the original images that detected as original and FN is the number of the manipulated images that detected as original falsely. In the pixel detection mode, these metrics are defined as the number of the true or manipulated images and their detection results.
 
 The proposed method has 5 adjustable parameters. Table 3 shows 4 parameter values for analyzing. The important and effective parameter in the detection result I the matching threshold in the key points matching step. Figure 4 shows average of the TPR and FPR values for different values of this parameter in several images from the small dataset. As it is shown, using threshold above 0.72 decrease the TP and increase the FP which is due to the increase of the false matching and inability of the method in estimating appropriate matrix. Moreover, using the threshold below 0.68 decreases the true positives. So in the proposed method, the 0.7 is used for the threshold. In the following of this paper, the proposed method is compared with the Amerini et. al. method [4] which their method has three parameters: the clustering method, the clustering cut off threshold and the minimum number of pixels in a cluster. These parameters are set based on the best represented results in the paper. These parameters are: the ward linkage clustering with cut off threshold of 2.2 and the minimum number of 3 pixels in a cluster.

 \begin {table}[H]
 \caption {parameters values} \label{tab:title} 
 \begin{center}
 \begin{tabular} {|c|c|p{3cm}|} \hline
   \centering{\textbf{Parameter}} & \textbf{Value} & \textbf{Description} \\ \rowcolor[gray]{.9}
         \hline
       C	& 5	& Number of clusters     \\ 
         \hline
       K	& 3	& Number of match able points for each key point \\  \rowcolor[gray]{.9}
         \hline
        P	& 2	& Matching coefficient \\
         \hline
       M & 	2	& Fuzzification coefficient \\ 
          \hline
 \end {tabular}
 \end{center}
 \end {table}

 \begin{figure}
\centering
\includegraphics[width=0.7\linewidth]{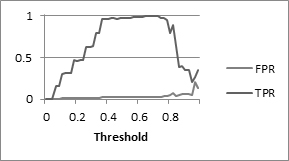}
\caption{TPR and FPR values for different thresholds}
\label{fig:4}
\end{figure}

\subsection{Key point level detection}
Choosing the appropriate matching threshold in the key points matching is important. The number and the distribution of the matched points is dependent on this threshold directly. In this part the proposed method and the Amerini method is compared for the key points detection. Figure 5 shows the number of detected matches by the proposed method and the Amerini method for two images (one simple forgery and one professional forgery). Figure 5 (a) is made by copying a rectangular area around the car. Figure 5 (b) shows the detection result of Amerini method and (c) shows the result of the proposed method. Number of matched key points for the proposed method is 89 key points while the Amerini method detects 70 key points. Figure 5 (d)- (f) is comparison of the proposed method and Amerini method for an image from professional forgery dataset. The Amerini method (e) just detects 48 matched key points and the proposed method detects 83 key points (nearly double). So the proposed method can detects more matched points. Moreover, these points are distributed along the whole copied area. This cause the proposed method to be able to estimate transformation matrix for the whole duplicated area and thus have a more precise detection. Furthermore, more matches leads better transformation matrix.

\begin{figure}
\centering
\includegraphics[width=0.9\linewidth]{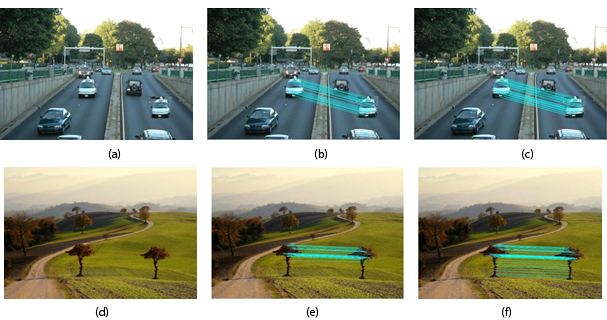}
\caption{Results in the key points level}
\label{fig:5}
\end{figure}

\subsection{detection for the dataset}
In this section three types of the images in the small dataset is chosen and their pixel level result is showed in the Figure 6. The first row of the figure 6 shows the manipulated images and the second row shows the detection results of the proposed method. The first column just manipulated by copy-move and rotation is applied on the second column and scaling is applied on the third column. The proposed method detects the duplicated area completely. Furthermore, the edges are detected with high precision.

\begin{figure}
\centering
\includegraphics[width=0.9\linewidth]{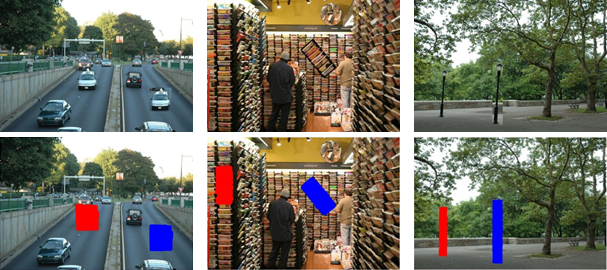}
\caption{Results of the proposed method on the simple forgery images}
\label{fig:6}
\end{figure}

Table 4 and 5 shows the result of the proposed method on the small dataset. Table 4 illustrates the image level detection of the proposed method and Amerini method. TPR in the proposed method and Amerini method are close to each other, and both methods detect 149 forgery images out of 150 But the FPR for the Amerini method is more than the proposed method. The Amerini method is biased to detect forgery while the proposed method detects forgery images with high precision and original images with low error. 

\begin {table}[H]
\caption {Image level detection results} \label{tab:title} 
\begin{center}
\begin{tabular} {|c|c|c|} \hline
Method	& TPR	& FPR \\
\hline
Amerini method	& 99.33	& 9.09 \\
\hline
Proposed method	& 99.33	& 6.36 \\
\hline
\end {tabular}
\end{center}
\end {table}

Table 5 shows the pixel level detection results of the proposed method. This table shows that the proposed method detects the manipulated areas with low error and high precision.

\begin {table}[H]
\caption {Pixel level detection result on the dataset} \label{tab:title} 
\begin{center}
\begin{tabular} {|c|c|c|} \hline
Method	& TPR	& FPR \\
\hline

Proposed method	& 92.84	& 2.73 \\
\hline
\end {tabular}
\end{center}
\end {table}

Table 4 shows the results of the proposed method and Amerini method on the 2000 image dataset. As it is shown, the proposed method has better precision and lower error rate comparing to the Amerini method. So the proposed method has better results in both datasets. 

\begin {table}[H]
\caption {Comparison results of the image level detection} \label{tab:title} 
\begin{center}
\begin{tabular} {|c|c|c|} \hline
Method	& TPR	& FPR \\
\hline
Amerini method	& 93.42	& 11.61 \\
\hline
Proposed method	& 93.71	& 8.38 \\
\hline
\end {tabular}
\end{center}
\end {table}

\subsection{key points matching improvement}
In this section the effect of using α matrix is analyzed. Figure 1 (a) shows the output of the key points matching where blue points are the key points level detection by the SIFT algorithm. Figure 7 (right) shows the key points level detection of the proposed method which the yellow points are the updated pixels. As it is shown in this figure, key points matches has been updated. For example the key point in the down of the figure 7(left) first matched wrongly, and during algorithm process it matched with another neighbor point, same as pixel in the top of the figure 7(left). There is another match in the top of the figure 7(left) which removed during the algorithm process because there is no appropriate key point near it and also its error was high.

\begin{figure}
\centering
\includegraphics[width=0.9\linewidth]{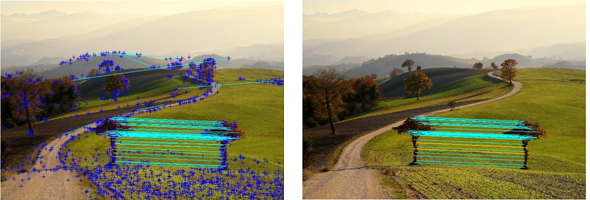}
\caption{Improving key points matching}
\label{fig:7}
\end{figure}

\subsection{Transformation matrix estimation}
Transformation matrix has a huge effect on the results of the copy-move forgery detection methods. Having a precisely estimated transformation matrix improves the detection accuracy. Table 7 shows the mean absolute error (MAE) of the transfer matrix, rotation and scaling parameters of the proposed method and the Amerini method. These results show that the proposed method calculate a more precise transformation matrix.

\begin {table}[H]
\caption {Mean absolute error of the parameters for the small dataset} \label{tab:title} 
\begin{center}
\begin{tabular} {|c|c|c|c|c|c|} \hline
Method	& $\theta$ & 	$S_x$ & 	$S_y$	& $t_x$	& $t_y$ \\
\hline
Amerini method	& 1.1828	& 0.0356	& 0.0375	& 2.0575	& 4.9822 \\
\hline
Proposed method	& 2.7994 & 	0.0195	& 0.0163 &	2.7994 & 	3.7657 \\
\hline
\end {tabular}
\end{center}
\end {table}

Tables 8-12 represent the precision of the transformation matrix estimation on the car image (Figure 2(a)) for different parameters and 15 attack scenarios (table 2)). In these tables 75 (15*5) comparison is made between the proposed method and the Amerini method and just in 12 cases the Amerini method is better than the proposed method and in the 63 cases the proposed method works better.

\begin {table}[H]
\caption {$t_x$ error on the car image with different attack scenarios} \label{tab:title} 
\begin{center}
\begin{tabular} {|p{1cm}|p{1cm}|p{1.3cm}|p{1cm}|p{1.3cm}|p{1cm}|} \hline
Method/ Attack	& True Value & 	Amerini Estimation & 	Amerini Error	& Proposed Method Estimation	& Proposed Method Error \\
\hline
A1 & 304 & 304.0969 & 0.0969 & 303.9407 & \textbf{0.0593} \\
\hline
A2 & 304 & 302.2825 & 1.7175 & 303.9007 & \textbf{0.0993} \\
\hline
A3 & 304 & 306.3537 & 2.3537 & 305.7881 & \textbf{1.7881} \\
\hline
A4 & 304 & 302.053 & 1.9467 & 305.3933 & \textbf{1.3933} \\
\hline
A5 & 304 & 301.3205 & 2.6795 & 305.1882 & \textbf{1.1882} \\
\hline
A6 & 304 & 302.1933 & 1.8087 & 305.0601 & \textbf{1.0601} \\
\hline
A7 & 304 & 302.2105 & \textbf{1.7895} & 301.9987 & 2.0013 \\
\hline
A8 & 304 & 301.9611 & 2.0389 & 305.8217 & \textbf{1.8217} \\
\hline
A9 & 304 & 305.9849 & 1.9849 & 305.9026 & \textbf{1.9026} \\
\hline
A10 & 304 & 301.4046 & 2.5954 & 302.1406 & \textbf{1.8594} \\
\hline
A11 & 304 & 305.8180 & 1.8180 & 305.8132 & \textbf{1.8132} \\
\hline
A12 & 304 & 306.4807 & 2.4708 & 302.8297 & \textbf{1.1703} \\
\hline
A13 & 304 & 301.5846 & 2.4154 & 304.1425 & \textbf{0.1425} \\
\hline
A14 & 304 & 301.3068 & 2.6932 & 303.3161 & \textbf{0.6839} \\
\hline
A15 & 304 & 301.3909 & 2.6091 & 305.8148 & \textbf{1.8148} \\
\hline
mean & - & - & 2.0685 & - & \textbf{1.2532} \\
\hline
\end {tabular}
\end{center}
\end {table}

\begin {table}[H]
\caption {$t_y$ error on the car image with different attack scenarios} \label{tab:title} 
\begin{center}
\begin{tabular} {|p{1cm}|p{1cm}|p{1.3cm}|p{1cm}|p{1.3cm}|p{1cm}|} \hline
Method/ Attack	& True Value & 	Amerini Estimation & 	Amerini Error	& Proposed Method Estimation	& Proposed Method Error \\
\hline
A1 & 81 & 80.9785 & 	0.0215	& 81.028	& \textbf{0.028}
 \\
\hline
A2 & 81 & 81.3241	& 0.3241	& 81.827	& \textbf{0.827}
 \\
\hline
A3 & 81 & 82.1902 & 	1.1902	& 80.0362	& \textbf{0.9638}
 \\
\hline
A4 & 81 & 78.406	& 2.594	& 79.9241	& \textbf{1.0759}
 \\
\hline
A5 & 81 & 78.6373	& 2.3627	& 81.134	& \textbf{0.134}
 \\
\hline
A6 & 81 & 78.8016	& \textbf{2.1984}	& 78.4188	& 2.5812
 \\
\hline
A7 & 81 & 78.9429	& 2.0571	& 79.9971	& \textbf{1.0029}
 \\
\hline
A8 & 81 & 79.3498	& \textbf{1.6502}	& 79.0635	& 1.9365
 \\
\hline
A9 & 81 & 82.7109	& 1.7109	& 79.7087	& \textbf{1.2913}
 \\
\hline
A10 & 81 & 79.4892	& 1.5108	& 79.625	& \textbf{1.375}
 \\
\hline
A11 & 81 & 82.9851	& 1.9851	& 82.8796	& \textbf{1.8796}
 \\
\hline
A12 & 81 & 83.0182	& 2.0182	& 81.4717	& \textbf{0.4717}
 \\
\hline
A13 & 81 & 79.5183	& 1.4817	& 81.6704	& \textbf{0.6704}
 \\
\hline
A14 & 81 & 78.3108	& 2.6892	& 79.9482	& \textbf{1.0518}
 \\
\hline
A15 & 81 & 83.2318	& 2.2318	& 82.3225	& \textbf{1.3225}
 \\
\hline
mean & - & - & 1.735
 & - & \textbf{1.1074
 } \\
\hline
\end {tabular}
\end{center}
\end {table}

\begin {table}[H]
\caption {$S_x$ error on the car image with different attack scenarios} \label{tab:title} 
\begin{center}
\begin{tabular} {|p{0.9cm}|p{1cm}|p{1.3cm}|p{1cm}|p{1.3cm}|p{1cm}|} \hline
Method/ Attack	& True Value & 	Amerini Estimation & 	Amerini Error	& Proposed Method Estimation	& Proposed Method Error \\
\hline
A1 & 1 & 	0.9957	& 0.0043	& 0.9998	& \textbf{0.0002}

 \\
\hline
A2 & 1	& 0.9976	& \textbf{0.0024}	& 0.9971	& 0.0029

 \\
\hline
A3 & 1	& 0.9961	& 0.0039	& 0.9998	& \textbf{0.0002}

 \\
\hline
A4 & 1	& 0.9989	& \textbf{0.0011}	& 0.9983	& 0.0017

 \\
\hline
A5 & 1	& 0.9972	& 0.0028	& 0.9985	& \textbf{0.0015}

 \\
\hline
A6 & 1	& 0.9992	& 0.0008	& 0.9999	& \textbf{0.0001}

 \\
\hline
A7 & 1.2	& 1.1987	& 0.0013	& 1.1996	& \textbf{0.0004}

 \\
\hline
A8 & 1.3	& 1.3005	& 0.0005	& 1.2998	& \textbf{0.0002}

 \\
\hline
A9 & 0.8	& 0.7934	& 0.0066	& 0.8021	& \textbf{0.0021}

 \\
\hline
A10 & 0.75	& 0.748	& 0.002	& 0.7497	& \textbf{0.0003}

 \\
\hline
A11 & 0.85	& 0.8371	& 0.0129	& 0.8485	& \textbf{0.0015}

 \\
\hline
A12 & 1.2	& 1.1931	& 0.0069	& 1.1997	& \textbf{0.0003}

 \\
\hline
A13 & 0.8	& 0.7872	& 0.0128	& 0.8044	& \textbf{0.0044}

 \\
\hline
A14 & 0.75	& 0.7525	& 0.0025	& 0.7507	& \textbf{0.0007}

 \\
\hline
A15 & 1.4	& 1.3976	& 0.0024	& 1.3986	& \textbf{0.0014}

 \\
\hline
mean & - & - & 0.0042
 & - & \textbf{0.0011} \\
\hline
\end {tabular}
\end{center}
\end {table}

\begin {table}[H]
\caption {$S_y$ error on the car image with different attack scenarios} \label{tab:title} 
\begin{center}
\begin{tabular} {|p{0.9cm}|p{1cm}|p{1.3cm}|p{1cm}|p{1.3cm}|p{1cm}|} \hline
Method/ Attack	& True Value & 	Amerini Estimation & 	Amerini Error	& Proposed Method Estimation	& Proposed Method Error \\
\hline
A1 & 1	& 0.9985	& 0.0015	& 0.9997	& \textbf{0.0003}

 \\
\hline
A2 & 1	& 0.9938	& 0.0062	& 0.9992	& \textbf{0.0008}

 \\
\hline
A3 & 1	& 0.9946	& 0.0054	& 0.9958	& \textbf{0.0042}

 \\
\hline
A4 & 1	& 0.9993	& \textbf{0.0007}	& 0.9972 & 	0.0028
 \\
\hline
A5 & 1	& 0.9994	& \textbf{0.0006}	& 0.9983	& 0.0017
 \\
\hline
A6 & 1	& 0.9995	& 0.0005	& 0.9996	& \textbf{0.0004}

 \\
\hline
A7 & 1.2	& 1.1985	& 0.0015	& 1.1992	& \textbf{0.0008}
 \\
\hline
A8 & 1.3	& 1.2993	& 0.0007	& 1.3002 & 	\textbf{0.0002}

 \\
\hline
A9 & 0.8	& 0.7987	& 0.0013	& 0.7994	& \textbf{0.0006}
 \\
\hline
A10 & 0.85	& 0.857	& 0.007	& 0.8487	& \textbf{0.0013}
 \\
\hline
A11 & 0.75	& 0.7493	& 0.0007	& 0.7499	& \textbf{0.0001}
 \\
\hline
A12 & 1.2	& 1.2027	& 0.0027	& 1.2003	& \textbf{0.0003}
 \\
\hline
A13 & 0.8	& 0.8095	& 0.0095	& 0.801	& \textbf{0.001}
 \\
\hline
A14 & 0.85	& 0.8278	& 0.0222	& 0.8488	& \textbf{0.0012}
 \\
\hline
A15 & 1.2	& 1.206	& 0.006	& 1.2064	& \textbf{0.0064}
 \\
\hline
mean & - & - & 0.0044
 & - & \textbf{0.0014 } \\
\hline
\end {tabular}
\end{center}
\end {table}

\begin {table}[H]
\caption {$\theta$ error on the car image with different attack scenarios} \label{tab:title} 
\begin{center}
\begin{tabular} {|p{0.9cm}|p{1cm}|p{1.3cm}|p{1cm}|p{1.3cm}|p{1cm}|} \hline
Method/ Attack	& True Value & 	Amerini Estimation & 	Amerini Error	& Proposed Method Estimation	& Proposed Method Error \\
\hline
A1 & 0	& 0.0577	& 0.0577	& 0.0048	& \textbf{0.0048}
 \\
\hline
A2 & 10	& 10.1723	& 0.1723	& 9.9833	& \textbf{0.0167}
 \\
\hline
A3 & 20	& 19.8115	& 0.1885	& 19.9038	& \textbf{0.0962}
 \\
\hline
A4 & 30	& 30.0218	& 0.0218	& 30.0121	& \textbf{0.0121}
 \\
\hline
A5 & 40	& 40.0228	& \textbf{0.0228} & 	39.9671 & \textbf{0.0329}
 \\
\hline
A6 & 50	& 49.6649	& 0.3351	& 50.007	& \textbf{0.007}
 \\
\hline
A7 & 0	& 0.0571	& 0.0571	& 0.016	& \textbf{0.016}
 \\
\hline
A8 & 0	& 0.0386	& 0.0386	& 0.0114	& \textbf{0.0114}
 \\
\hline
A9 & 0	& 0.0933	& 0.0933	& 0.0097	& \textbf{0.0097}
 \\
\hline
A10 & 0	& 0.116	& 0.116	& 0.0903	& \textbf{0.0903}
 \\
\hline
A11 & 0	& 0.1878	& 0.1878	& 0.0489	& \textbf{0.0489}
 \\
\hline
A12 & 30	& 50.0538	& 0.0538	& 30.0107	& \textbf{0.0107}
 \\
\hline
A13 & 30	& 30.3507	& 0.3507	& 29.9379	& \textbf{0.0621}
 \\
\hline
A14 & 35	& 35.1407	& 0.1407	& 34.8763	& \textbf{0.1237}
 \\
\hline
A15 & 35	& 36.0746	& 1.0749	& 34.9777	& \textbf{0.0223}
 \\
\hline
mean & - & - & 0.194 & - & \textbf{0.0376} \\
\hline
\end {tabular}
\end{center}
\end {table}

\subsection{professional image forgery detection}
The manipulated images in the previous sections made by duplicating a square or rectangular area. But in the real world there are strong image editing programs such as Photoshop, Image Doctor which help to generate duplicated regions with high sensitivity and elegance. Furthermore, Simple forgery images can be easily detected by the eye and detecting them does not need detection algorithm. On the other side, professional forgery images are not detectable by eye and using a detection algorithm is needed to analyze them. These images can be a good metric to analyze the efficiency of the algorithms. To analyze the proposed algorithm we test the algorithm using few professional forgery images that we generate using the Photoshop software.

Figure 8 shows the performance of the proposed method in cases of rotation, scaling up and down and their combination. The Figure 8(a) is the original image which all the manipulation is done on this image to hide the weasel and its around area. Figure 8(b) shows the manipulated image with the rotation of 50 degree and 8(c) shows the detection result. As it is shown even the tale of the weasel which covered by a thin area is detected. 

Figure 8(d) shows the scale up manipulation with the scale of 1.25 and 8(e) shows the detection result. In this manipulation, the original area scale up 1.25 and located in the target area. However in this manipulation, the original area is small (especially the weasel tale) but it detected precisely. 8(f) is generated using scale down with 0.76 coefficient. The detection result is shown in 8(g).

Figure 8(h) is generated using combination of rotation and scaling. The rotation is 50 degree and the scaling is 0.75. The detection result is shown in 8(i) which the proposed method detects even the small areas such as tale, foot and head of the weasel.

\begin{figure}
\centering
\includegraphics[width=0.9\linewidth]{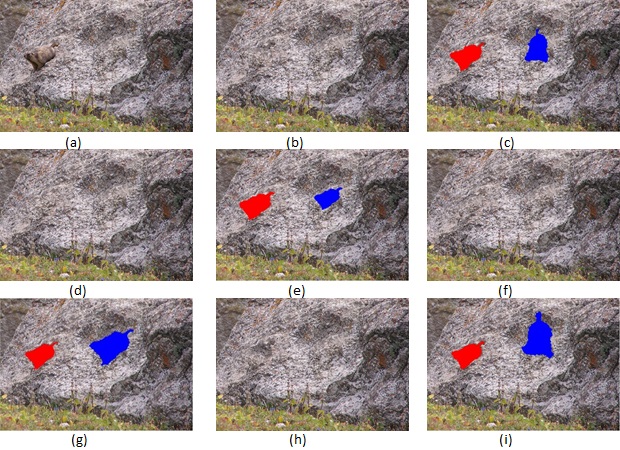}
\caption{Results of the proposed method in case of rotation, scaling and their combination}
\label{fig:8}
\end{figure}

In the figure 9, some challenging images in the field of copy-move forgery detection is shown to analyze the proposed method. In this figure, the original images, manipulated images and the detection results of the proposed method is shown. In all cases, the proposed method detects the copied area with all details even if this area was small.

In the basket image (figure 9(a)), the edges are detected with high precision. In the wall image (figure 9(d)), the proposed method detects the copied area with high precision and in the tree image (figure 9(g)) shows the strength of the proposed method which detects the thin trunk of the tree precisely.

\begin{figure}
\centering
\includegraphics[width=0.9\linewidth]{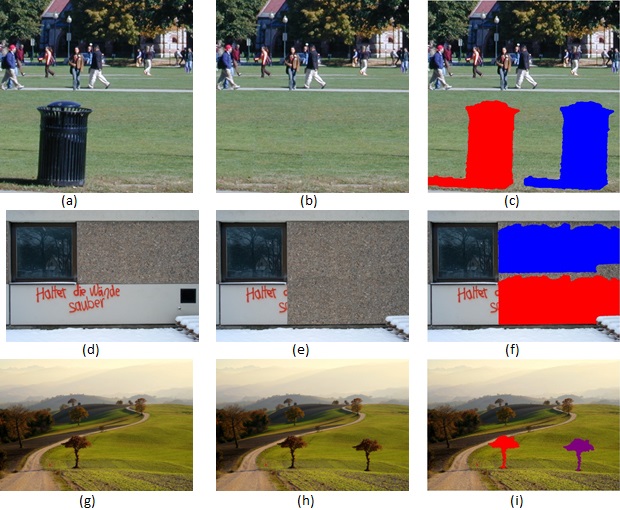}
\caption{Evaluation of the proposed method with the professional forgery images}
\label{fig:9}
\end{figure}

\subsection{detecting multiple copied areas}
There are two ways for copying multiple area in the image: some manipulation in one image and copying one area multiple time. In the first case, some independent manipulation is done on the image. Figure 10(b) shows this kind of manipulation which has done on the figure 10(a). In this image, two birds are covered with two independent areas. The algorithm needs to be executed multiple times to detect these kinds of manipulation, in the previous methods such as [31]. But in the proposed method, since the transformation matrix is estimated locally, the algorithm can detect all copied areas in one execution. The result of the proposed method shown in the figure 10(c) and it can be seen that the proposed method even detects the foot, and wings of the birds.

In some cases, it is possible to copy one area in some areas of the original image and the manipulated image is generated. The goal of this manipulation is to hide some objects by copying just one area. Figure 10(d) shows one sample of this manipulation and the detection result is shown in figure 10(e). The proposed method detects the copied area precisely in this case too. Since the g2NN metric is used in the proposed method, it can detect all copied areas just in one execution.

\begin{figure}
\centering
\includegraphics[width=0.9\linewidth]{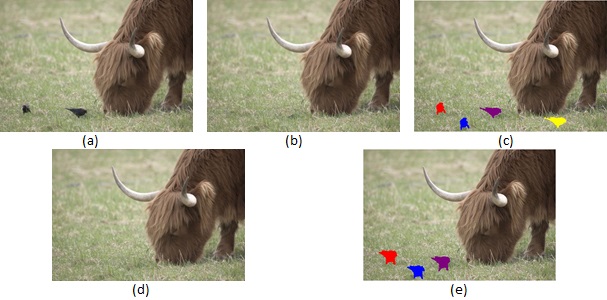}
\caption{Result of detecting multiple copied zones}
\label{fig:10}
\end{figure}

\section{Conclusion}
In this paper a cost function is defined to have interaction between different steps of the copy-move forgery detection methods based on the key point. This cost function includes matching, clustering and transform estimation steps. The proposed method calculated the clustering, transformation, key points matching matrixes and cluster centers by an iterative routine such that the cost function is minimized. The experimental results show that the proposed method can detect different kinds of copy-move image forgery more precisely than the previous methods in just one execution.

Geometric error is used to calculate the transformation matrix error in the proposed method. To continue this research, other methods can be used to calculate the error and the distance. Moreover, using inverse transformation matrix may lead to estimate the transformation matrix with lower error, since the error of the transformation matrix and its inverse are not the same at a point. The proposed method can be applied on the block based methods and instead of key points, using block centers is possible. So all the matrixes must be defined based on the blocks.

\ifCLASSOPTIONcaptionsoff
  \newpage
\fi



%
\bibliographystyle{unsrt}
\bibliography{bare_conf2}




\end{document}